# Effect of Oxygen Defects Blocking Barriers on Gadolinium Doped Ceria (GDC) Electro-Chemo-Mechanical Properties

Ahsanul Kabir[1,3*], Simone Santucci[1], Ngo Van Nong[1], Maxim Varenik[2], Igor Lubomirsky[2], Robin Nigon[3], Paul Muralt[3], Vincenzo Esposito[1*]

[1]Department of Energy Conversion and Storage, Technical University of Denmark, Frederiksborgvej 399, Roskilde 4000, Denmark

[2]Department of Materials and Interfaces, Weizmann Institute of Science, Rehovot 76100, Israel

[3]Ceramics Laboratory, Dept. of Materials, Swiss Federal Institute of Technology, Lausanne 1015, Switzerland

*Corresponding Authors: E-mail: ahsk@dtu.dk, vies@dtu.dk

## Abstract

Some oxygen defective metal oxides, such as cerium and bismuth oxides, have recently shown exceptional electrostrictive properties that are even superior to the best performing lead-based electrostrictors, *e.g.* lead-magnesium-niobates (PMN). Compared to piezoelectric ceramics, electromechanical mechanisms of such materials do not depend on crystalline symmetry, but on the concentration of oxygen vacancy ($V_O^{\cdot\cdot}$) in the lattice. In this work, we investigate for the first time the role of oxygen defect configuration on the electro-chemo-mechanical properties. This is achieved by tuning the oxygen defects blocking barrier density in polycrystalline gadolinium doped ceria with known oxygen vacancy concentration, $Ce_{0.9}Gd_{0.1}O_{2-\delta}$, $\delta$ = 0.05. Nanometric starting powders of *ca.* 12 nm are sintered in different conditions, including field assisted spark plasma sintering (SPS), fast firing and conventional method at high temperatures. These approaches allow controlling grain size and Gd-dopant diffusion, *i.e.* via thermally driven solute drag mechanism. By correlating the electro-chemo-mechanical properties, we show that oxygen vacancy distribution in the materials play a key role in ceria electrostriction, overcoming the expected contributions from grain size and dopant concentration.

Keywords: Electrostriction, Vacancies, Gadolinium doped ceria, Sintering

# 1. Introduction

Cerium oxide ($CeO_2$) has been comprehensively investigated in last few decades due to its multifold applications, more specifically in electro-ceramics and catalysts [1]–[4]. It has a centrosymmetric fluorite structure with a pronounced oxygen defectivity, *i.e.* oxygen vacancies ($V_O^{\cdot\cdot}$). This feature makes ceria an excellent ionic conductor, especially suitable for solid-state electrolytes at high temperatures [5], where acceptor dopants are used to enhance oxygen defects concentration [$V_O^{\cdot\cdot}$], in the lattice [6][7]. In cerium oxide, $Ce^{4+}$ cation can also be reduced to $Ce^{3+}$ under low oxygen partial pressure ($P_{O_2}$) at high temperatures, creating both *quasi*-free localized electrons, *i.e.* small polarons and oxygen vacancies [8], resulting mixed ionic electronic conductivity (MIEC) [5][9]. Beside these properties, ceria exhibits non-classical giant electrostriction properties at room temperatures [10]–[12], both in thin films and bulk materials. Remarkably, ceria thin film expands perpendicular to applied field direction, with a large compressive stress (≈ 500 MPa) [10]. The average electrostriction coefficient ($M_e$) is reported as ≈ $6.5 \cdot 10^{-18}$ $(m/V)^2$ for [$V_O^{\cdot\cdot}$] = 5%, *i.e.* 20 mol% Gd-doped ceria [10]. Such value is high for a material with low dielectric constant ($\varepsilon_r^{GDC} \approx 30$) [13], even higher compared to relaxor ferroelectric metal oxides, *e.g*. Ca-doped PMN (Pb,Mg)$NbO_3$ ($\varepsilon_r^{Ca\text{-}PMN} \approx 4000$) [14]. Yavo *et al.* also verified this type of electromechanical properties in bulk gadolinium doped ceria and another oxygen defective fluorite oxide ($Bi_2O_3$), which exhibit similar results, thus representing to a new class of electroactive materials [12][15]. The atomistic model proposed by Lubomirsky *et al.* based on XANES/EXAFS measurements [16] comprehensively explains the underlying phenomena of this unusual behavior, further suggested that the presence of oxygen vacancy makes distorted $Ce_{Ce}$-$7O_O$-$V_O^{\cdot\cdot}$ units: consisting of contracted Ce-O and expanded Ce-$V_O^{\cdot\cdot}$ bond, compared to Ce-O bond in Ce-8O unit [10][16]. As a result, asymmetric charge distribution and anisotropic local dipolar elastic field are developed in the fluorite lattice [17]. Under applied electric field, distorted $Ce_{Ce}$-$7O_O$-$V_O^{\cdot\cdot}$ complexes conform to a more fluorite like structure, subsequently local atomic displacement produces giant electromechanical effect [16]. Despite intriguing, some questions about the role of oxygen vacancies and microstructure on electrostriction still remains unexplored [12]. Besides, Lubomirsky and co-workers suggested a power law dependence of *I-V* relationship in grain boundary blocking behavior based on space charge mechanisms [18][19] and demonstrated that an increase of grain boundary resistance leads to a decrease in the portion of applied voltage drop in the bulk, decreasing electromechanical properties [20].

As ionic conductor at high temperatures, bulk properties of ceria based compounds are controlled by process parameters *i.e.* morphology of initial powders, sintering kinetics/thermal history, densification, final microstructure *etc*. For sintering and consolidation, mass diffusion mechanisms are especially dominated by solute drag phenomena, which, depends on both dopant size and valence. These can influence ionic configurations at the grain boundary, for instance, by trapping vacancies in disorder and/or in vacancies-ions complexes with low mobility [21][22]. This is described by the so called “brick model” that is observed for highly defective ceria where fast ionic migration mechanisms are activated, thus reducing cations trapping effects at the grain boundaries [23]–[26]. Solute drag phenomena creates specific grain boundary configurations and non-stoichiometry, which acts as blocking barriers to migrate charge species in the material, significantly affect intrinsic properties [21]. Moreover, Shibata *et al.* experimentally showed that the long-range electric interaction is the governing factor in controlling the local charge distribution at the crystal interface [27].

Based on previously published reports, designing the microstructure at nano-scale is also expected to create more significant differences between oxygen migration effects, revealing dissimilar physical and chemical properties than grain of micron sizes. In addition, decreasing the grain size leads to increase grain boundary effect on the material, as well as increasing the density of the blocking barrier [28]. On the other hand, Esposito *et al.* proposed that grain boundary blocking factor is not necessarily a geometrical factor [3]: at a fixed oxygen vacancy concentration, different grain boundary blocking effects are encountered, depending on the entity of solute drag effect, controlled by sintering conditions [3]. The grain boundary blocking effect is explained both theoretically and experimentally by the distribution of defects by space charge layer model [29]–[35]. Other techniques also can disrupt the solute drag effect, even maintaining the polycrystalline in the nanoscale. This occurs by field assisted sintering techniques (FASTs), such as spark plasma sintering (SPS) [36]. FASTs uses three important parameters (i) pulsed electric field, (ii) high heating rates and (iii) high pressure to preserve ultra-fine grains [37][38].

In the present work, we use nanometric 10 mol % gadolinium doped ceria (GDC10) to produce polycrystalline samples with different oxygen vacancy configuration. This is done by sintering the nano-powders by field assisted (SPS), fast firing and conventional method, which yields dense polycrystalline samples with tuned oxygen ions blocking barriers. A commercial high-density tape cast sample with minimized grain boundary is also used for comparison. The influence of oxygen vacancy configuration on electro-chemo-mechanical properties of GDC was investigated, comparing the electrochemical properties from low to intermediate temperatures (*ca*. 300-575 °C) with the electromechanical properties at room temperatures, expecting unchanged oxygen vacancy configuration in the materials.

# 2. Experimental Procedure

## 2.1 Powder Synthesis

Nano size gadolinium doped ceria GDC10 powders were prepared by co-precipitation method using diamine in aqueous solution [3]. Cerium nitrate hexahydrate (Sigma-Aldrich, USA) and gadolinium nitrate hexahydrate (Sigma-Aldrich, USA) salts were mixed together in stoichiometric proportions to prepare 0.1 M solution in deionized water. Then MDEA (N-methyl-diethanolamine) was added dropwise. The molar ratio between total cations and MDEA was 1:3. The resulting precipitates were kept overnight under mild stirring. Afterwards, the precipitates were centrifuged and washed several times with ethanol. The resulting gel was dried at 120 °C followed by calcination at 500 °C for 2 hours. After the calcination, hard agglomerated powders were ball-milled in ethanol with 2 mm zirconia balls for 10-12 hours at 50 rpm, followed by drying at 120 °C for 10 hours. Finally, the powders were softly crushed by mortar and pestle and sieved using a 150 µm mesh.

## 2.2 Pellet Preparation

The SPS sample was consolidated by field assisted spark plasma sintering (SPS) (Dr. Sinter Lab 515S, Japan) under high vacuum (≤ $6 \cdot 10^{-6}$ Torr) at 980 °C, uniaxial pressure of 70 MPa with 5 min dwelling. To minimize the chemical reduction that may occur by FAST treatments, the sample was re-oxidized by post-heating at 700 °C for 1 hour. For the conventional sample, powders were uniaxially cold pressed at 200 MPa for 30 s, followed by sintering at 1450 °C in air for 10 hours. To achieve high density in the pellets independently by the powders packing [39], the fast fired sample was pre-densified by SPS at 900 °C and then thermally treated at 1450 °C for 0.1 hour with 20 °C/min heating and cooling rate [3]. Commercial tape (Kerafol Germany) was sintered at 1450 °C for 2.5 hours.

## 2.3 Materials Characterization

Density of the samples was measured in water using Archimedes method. The particle size and phase composition of the samples were analyzed by transmission electron microscope (TEM) (JEOL 2100, USA) and X-ray diffraction technique (XRD) (Bruker D8, Germany) respectively. The microstructure was investigated by a high-resolution scanning electron microscope (SEM) (Zeiss Merlin, Germany). The grain sizes were calculated by the linear intercept method using a minimum of 100 grains, multiplying with correction factor 1.57 [40]. The electrochemical impedance spectroscopy (EIS) was performed at 300-575 °C in air using Solarton 1260 (UK), in a frequency range of 0.01 Hz to 1 MHz with a 100 mV alternate signal. The samples have bar-like geometry. Gold-silver mixture electrodes pastes were coated on top of the sample and dried at 600 °C for 15 min. Symmetric configuration using gold as electrodes, silver as current collectors and platinum wire as current leads were used. The EIS data were plotted using Real $Z'$ and Imaginary $Z''$ of the impedance normalized by the geometrical cell parameter $k$ of each sample, where $k = A/t$, $A$ is the electrode area and $t$ is the thickness. The resulting geometry normalized EIS plots are thus expressed as geometrically normalized Nyquist plots, *i.e.* ρ' vs ρ" (Cole-Cole plots), and as ρ" vs frequency plots (similar to Bode-plot). The latter related the relative dielectric constant, $\varepsilon_r = \frac{1}{-2 \cdot \pi \cdot f \cdot \rho''}$, as a function of the AC electric field frequency (f). The data were fitted by equivalent circuit and analyzed by ZView software shareware version. The electromechanical measurement was performed using a proximity sensor (Capacitance, Lion) based system with a lock in detection, as previously reported in [12][15]. Prior to measurement, the system was calibrated with PZT (Shenzhen Yuije Electronics Co. Ltd. China). The sample is pressed between two metal electrodes using a spring. A pushrod is used to transfer displacement from the electrodes to a proximity sensor. The signal from the proximity sensor is captured using a lock-in amplifier. Longitudinal electrostrictive strain (parallel to the applied electric field) is calculated as a ratio between the displacement and the original thickness of the ceramic pellets.

## 3. Results and Discussion

Use of nano-powder in ceramic processing allows a fine control of the microstructural features in final bulk materials. The morphology and structure of the starting nano-powder used in this work is shown in Fig.1. TEM analysis revealed that particles have spherical shape and are loosely agglomerated. The nano-powders have a narrow range of size distribution with an average particle size ranging between 10-15 nm. Electron diffraction pattern shows fluorite symmetry of ceria. Crystallography was further confirmed by X-Ray diffraction technique.

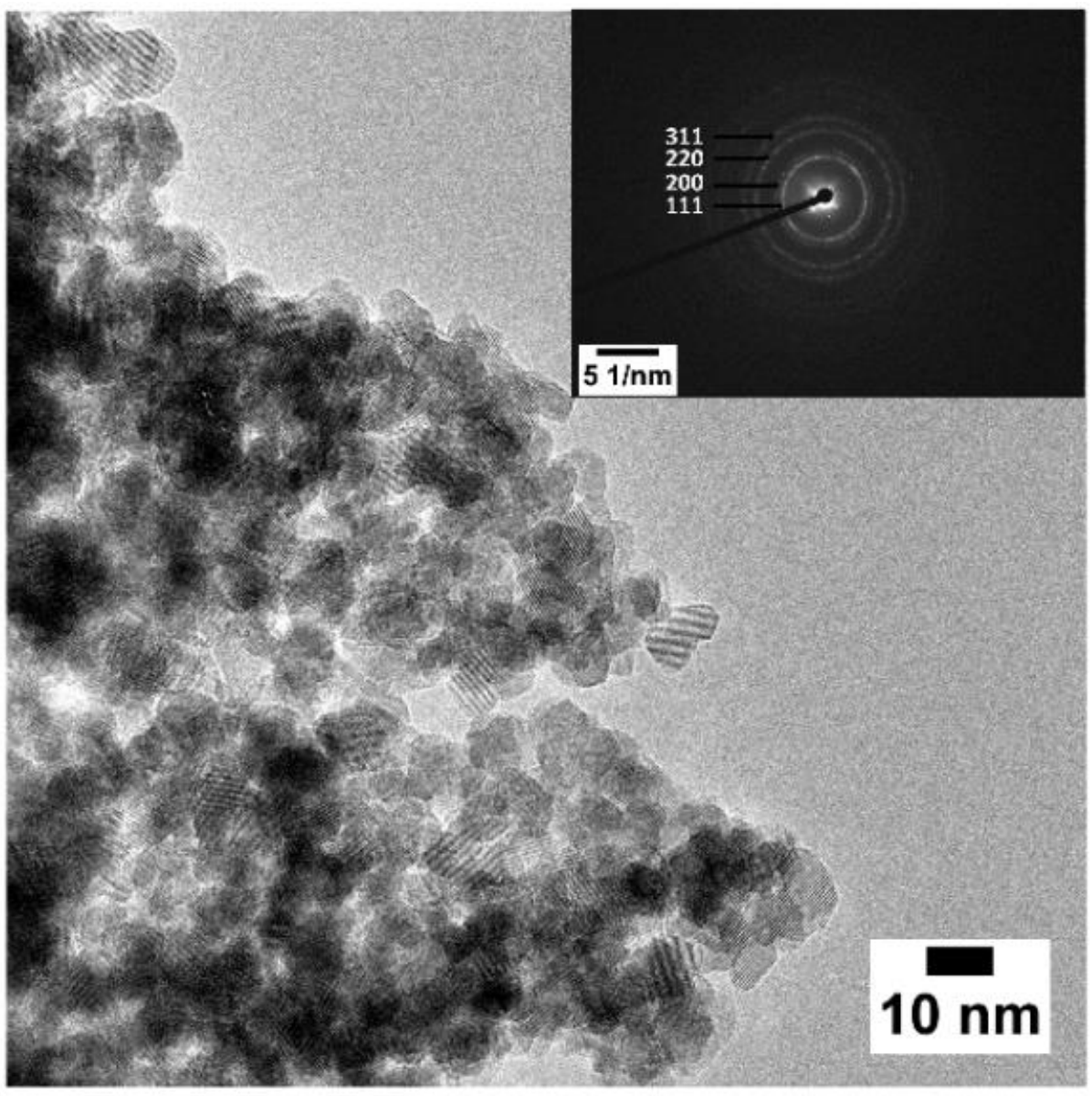

Figure 1: Bright field TEM images of GDC powder, calcined at 500 °C for 2 hours. Inset shows selective area electron diffraction (SAED) pattern of it.

Fig.2 illustrates the XRD pattern of GDC-powder and sintered samples. Within the detection threshold of the technique, the patterns reveal no secondary phases. The reflection peaks of the pattern perfectly fit with theoretical pattern (ICSD code 251473). Average crystallite size by Scherrer and lattice parameter for the starting powders are estimated as ≈ 12 nm and 0.540 nm respectively. Results are consistent with the TEM analysis in Fig.1. For the sintered samples, XRD patterns in Fig.2 also display identical results. Narrow peaks imply an increase of particle size during sintering, according to the Scherrer formula [41].

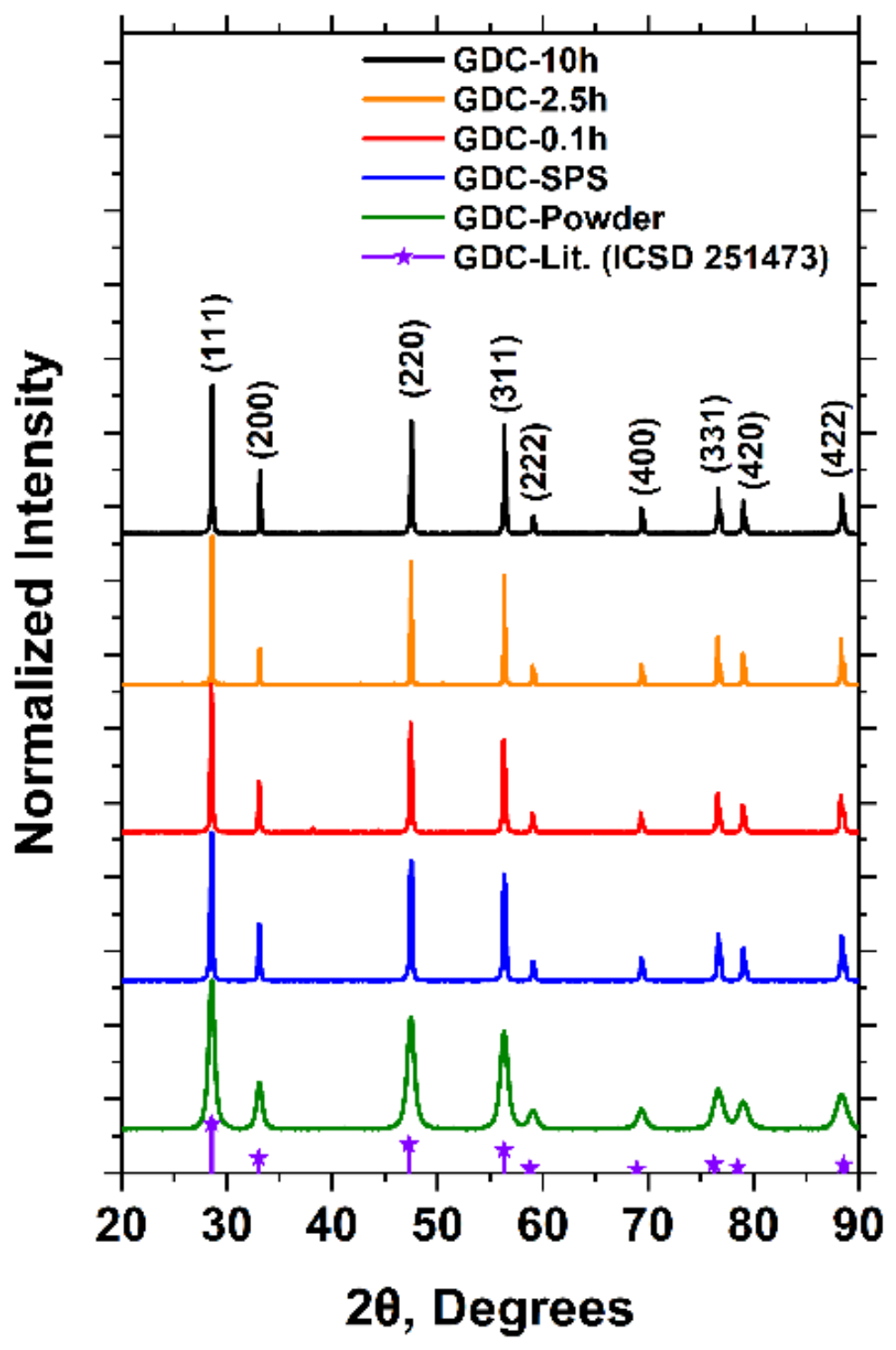

Figure 2: XRD pattern of gadolinium doped ceria (GDC) powder and sintered pellets.

The microstructures of sintered GDC samples are presented in Fig.3. The micrographs indicate that grains are highly dense with negligible intra-granular porosity. The outcome is consistent with experimental density of the pellets, which is above ≥ 96%, for all samples. Grain size analysis shows that SPS and fast firing sample has significantly smaller grain size (around 150-200 nm) than conventionally sintered materials. They exhibit typical polygonal grains with nearly homogeneous size distribution. Furthermore, they show no surface relaxation at the grain boundary. Plapcianu *et al.* found similar results in SPS sintering of GDC [37]. These authors stated that restricted grain growth in this type of non-conventional sintering is attributed to fast heating rates especially in the initial stage of sintering, where grain-coarsening mechanism dominates. However, both GDC-10h and GDC-2.5h materials show high degree of grain growth with grain size about 2.0 ± 0.3 µm and 1.5 ± 0.2 µm respectively. Nearly all grains have equilibrium shape at the triple point (red lines in Fig. 3.c, 3.d) with fully relaxed and residual small grain boundary curvature (see black arrows in Fig 3.c, 3.d).

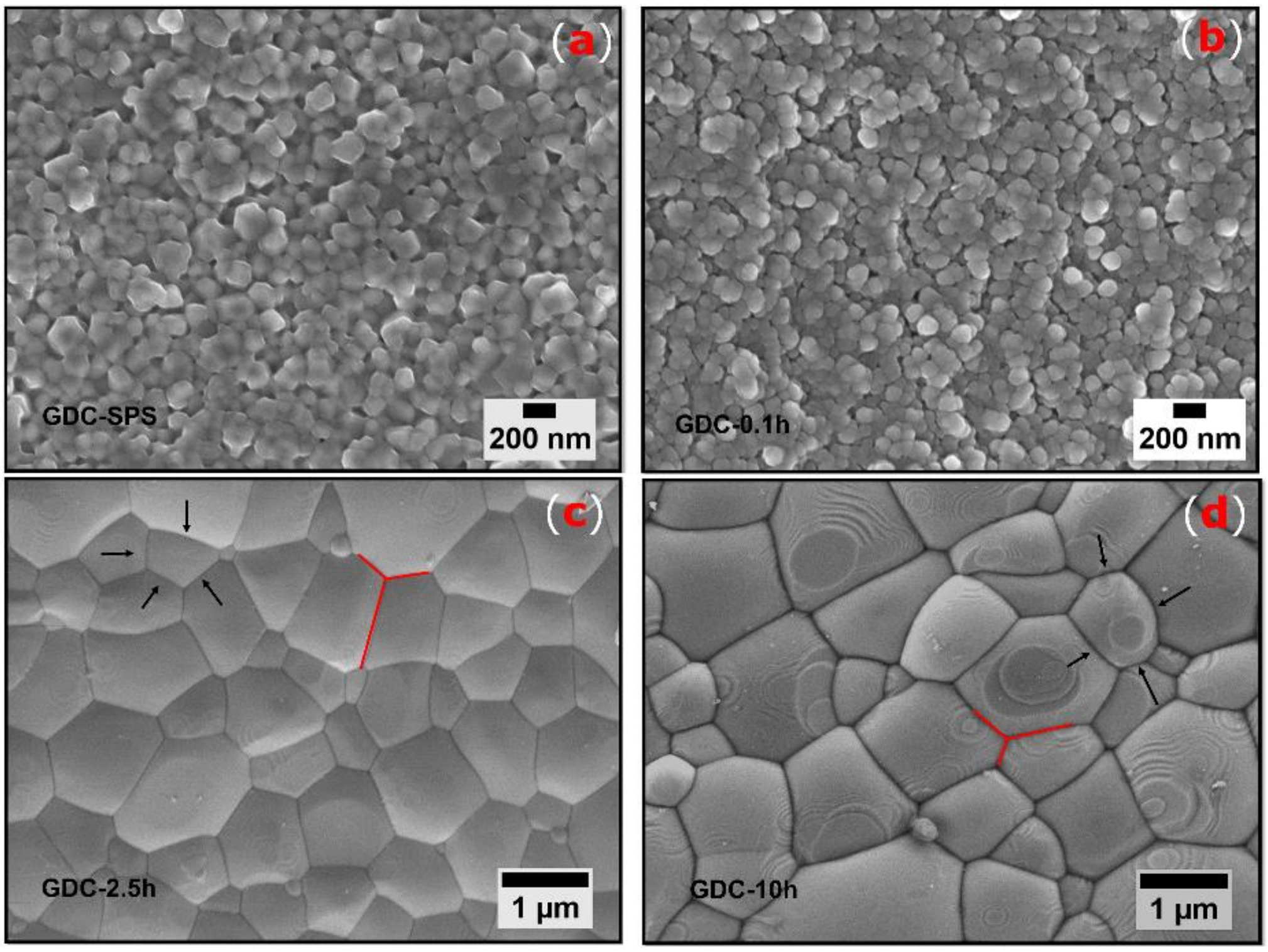

Figure 3: SEM images of four different GDC samples, sintered in (a) SPS at 980 °C, 70 MPa (b) air at 1450 °C, for 0.1h (c) air at 1450 °C, for 2.5h (d) air at 1450 °C, for 10h.

Fig.4 illustrates the geometry normalized Nyquist plots (ρ' vs ρ") at 300 °C. At this temperature, as evident from Fig.4.a, only one semicircle is observed in GDC-SPS and GDC-0.1h samples, which corresponds to overlapped bulk and grain boundary contribution. On the other hand, both the GDC-10h and GDC-2.5h samples display two semi-circle that refers to high and intermediate frequency associated bulk and grain boundary impedance respectively [42]. The low frequency arc attributes to the electrode/material interface polarization mechanisms that are not relevant for this discussion [43]. Despite having similar density and same dopant concentration, all the materials develop quite different total resistance/blocking barrier as seen in Fig.4.b, an effect that is ascribed to dissimilar oxygen vacancy migration mechanism. In a broader sense, various sintering mechanisms lead to govern unlike oxygen vacancy configurations, as well as dissimilar ordering of vacancies and defect hopping probabilities. Both GDC-10h and GDC-2.5h samples exhibit a comparable bulk resistance, however distinct grain boundary resistance are shown. Grain boundary blocking factor ($\alpha_{gb}$) is estimated as ≈ 0.65 and ≈ 0.9 for GDC-2.5h and GDC-10h respectively. Comparing the microstructure, such behavior is unexpected as GDC-10h has larger grain size, *i.e.* lower gb/grain geometrical ratio, than GDC-2.5h sample. Similar results were previously reported by Esposito *et al*., where long-term thermal treatment, *i.e.* 36 hours, possesses a detrimental effect on the blocking factors for low dopant concentration, *e.g.* for the conventional sintering for 10 hours [3]. Conversely, significantly higher resistance in GDC-SPS compared to GDC-10h indicates a field assisted trapping of oxygen vacancies in the lattice during SPS. High electric fields in SPS can result a frozen non-equilibrium dopant distribution, which could decrease the possible vacancy mobility in the nanometric polycrystalline materials. Moreover, some reports suggest that the chemical reducing condition occurring during the SPS, can also create large number of $Ce^{3+}$ species that remain confined at the grain boundary, even after long time re-oxidation [36]. As a result, grain boundary space charge potential would increase. Nanostructures in SPS create a high density of grain boundaries, which act as a high blocking barrier to charge migration [44][45]. Despite having similar nanostructure like GDC-SPS, fast firing sample, *i.e.* GDC-0.1h, presents an intermediary behavior between GDC-SPS and GDC-10h. Moreover, different values of resistance are attributed to the nature of solute drag effect due to various thermal treatments. Besides, ρ" vs frequency plots (Fig.4.b) elucidate the distribution of charge transport by means of its relaxation frequency. In Fig.4b, at 300 °C, both GDC-10h and GDC-2.5h expose a bulk relaxation frequency around ~100 kHz, while, grain boundary relaxation frequency exhibits around 1.5 kHz and 20 kHz, respectively. Such a low grain boundary frequency response is attributed to high blocking barrier effect in GDC-10h compound [3]. Furthermore, both the GDC-SPS and GDC-0.1h show an overlapped relaxation frequency at 1.2 kHz and 5 kHz respectively. This overlay behavior is ascribed to different charge transport mechanism, as resulting from unrelaxed microstructure and non-equilibrium fast thermal treatment.

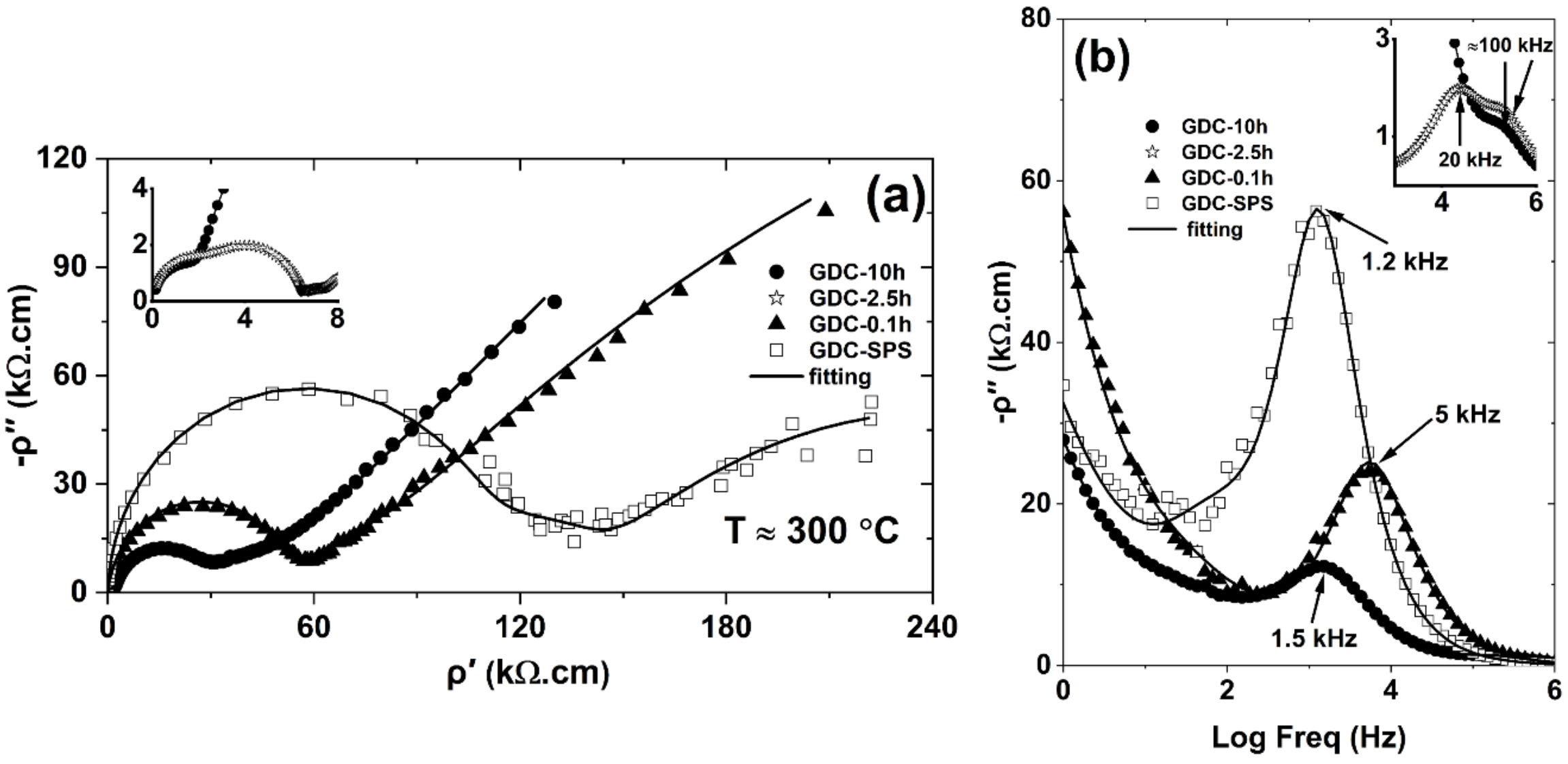

Figure 4: Geometry normalized (a) Nyquist plots (ρ' vs ρ") and (b) imaginary ρ" vs log frequency (f) plots of different GDC samples, at 300 °C. The Inset graphs compare GDC-2.5h sample with GDC-10h.

Table I summarizes the capacitance, relaxation frequency values of the samples at 300 °C, along with average grain size. These values are calculated from the constant phase element with the 1/RC relationship from the fitted plots.

Table 1: Grain size, capacitance and relaxation frequency response of the samples.

| Sample ID | $C_{total}$ (F) | | $f_{total}$ (Hz) | | Grain Size (nm) |
|---|---|---|---|---|---|
| GDC-SPS | $2.0 \times 10^{-11}$ | | $1.2 \times 10^{3}$ | | 200 ± 25 |
| GDC-0.1h | $3.5 \times 10^{-11}$ | | $5.0 \times 10^{3}$ | | 170 ± 20 |
| | $C_{Bulk}$ (F) | $C_{G.B.}$ (F) | $f_{Bulk}$ (Hz) | $f_{G.B.}$ (Hz) | Grain Size (µm) |
| GDC-10h | $1.2 \times 10^{-11}$ | $3.0 \times 10^{-10}$ | $1.0 \times 10^{5}$ | $1.5 \times 10^{3}$ | 2.0 ± 0.3 |
| GDC-2.5h | $2.9 \times 10^{-11}$ | $4.0 \times 10^{-10}$ | $1.3 \times 10^{5}$ | $2.1 \times 10^{4}$ | 1.5 ± 0.2 |

The temperature dependence of the total electrical conductivities ($\sigma = 1/\rho'$), *i.e.* bulk plus grain boundary is illustrated with an Arrhenius plot in Fig.5. As observed, conventionally sintered samples display superior electrical conductivity than non conventionals. GDC-2.5h displays highest electrical conductivity among all of them, whereas, GDC-SPS reveals the minimum, which is directly interlinked with activation energy values. The high activation energy in GDC-SPS is also consistent with the existence of large density of blocking barrier illustrated in Fig.4, an effect conceivably caused by vacancy trapping or/and vacancy clustering mechanisms.

The minimum activation energy observed in GDC-2.5h could be of different grain boundary composition (uniform dopant distribution) due to short thermal treament. GDC-0.1h shows similar conductivity of previous work of Esposito *et al.* (see Fig.5), however, as expected, GDC-10h sample displays much higher conductivity than GDC-36h [3].

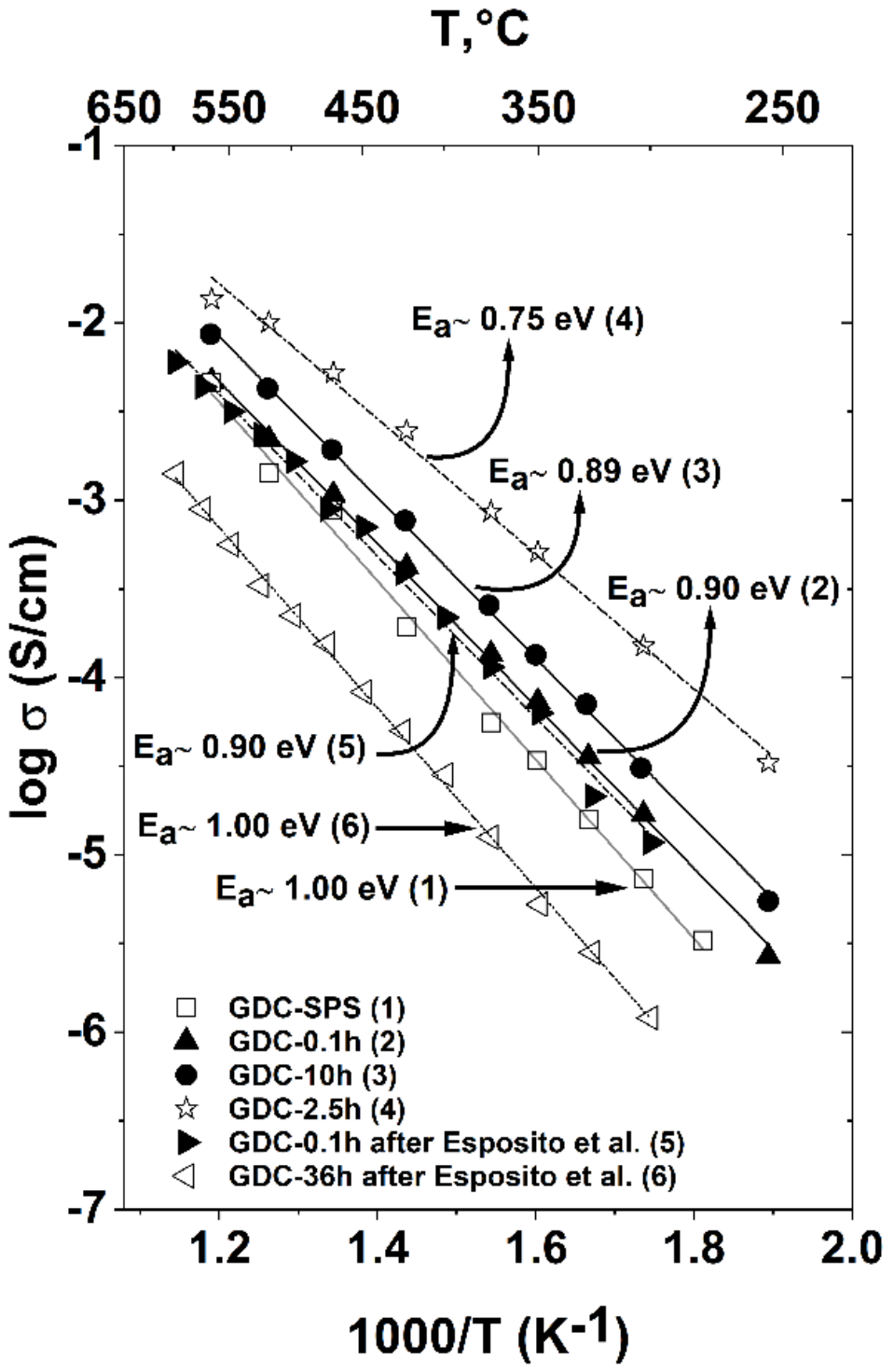

Figure 5: Arrhenius plot for the estimation of total electrical conductivities of the samples, measured in air. Results are compared with literature [3].

The electrostrictive strain as a function of applied electric field square is presented in Fig.6. All the compounds exhibit negative longitudinal strain that agrees with previous reports of GDC thin films and bulk materials [10][12]. They respond at second harmonic of the applied electric field with different frequency within the applied range, further confirming its electrostriction behavior. Besides, the graph explains the following trends *i.e.* at low frequency, the strain saturates with increasing electric field amplitude, whereas with increasing frequency magnitude of strain value declines dramatically. The strain saturation behavior empirically fits to the following equation:

$$u\,(E^2) = M_{33} \cdot E_{sat}^2 \cdot \left[1 - \exp(-E^2/E_{sat}^2)\right] \tag{1}$$

Where, $M_{33}$ is the electrostriction 3-3 strain coefficient, $E_{sat}$ is saturation electric field. Beyond the saturation point, linear relationship between electrostriction strain vs $E^2$ is no longer valid. As expected, GDC-SPS sample responds at much higher electric field compared to others, for instance, electrostrictive strain coefficient ($M_{33}$) value being one order of magnitude lower than GDC-10h. The small value of the former is not due to its unrelaxed grain size, but connected to trapped defects-cation association and its interaction with the electric field. These defect-complexes are neutral and do not respond at low field, which can be interrelated with its high density of blocking barriers. In contrast, GDC-10h generates high electrostrictive strain compared to both GDC-SPS and GDC-0.1h, further confirms the effect of blocking barrier in electrostriction. Surprisingly, GDC-2.5h, with low blocking barriers, shows $M_{33}$ value much smaller than GDC-10h. Furthermore, it does not show any strain saturation behavior, and strain linearly increases with $E^2$. The reason for this effect could be of high conductivity, which leads to marginal voltage drop at bulk grain. This significantly highlights the flexibility of blocking barrier in tuning the electrostrictive strain. For high electrostriction, the barriers should block oxygen vacancy migration without meaningfully decreasing the potential drop and vacancy should resonate within the lattice, as per the model described by Lubomirsky *et al.* [16]. Therefore, sample with a high resistive grain boundary would require high electric field to generate strain.

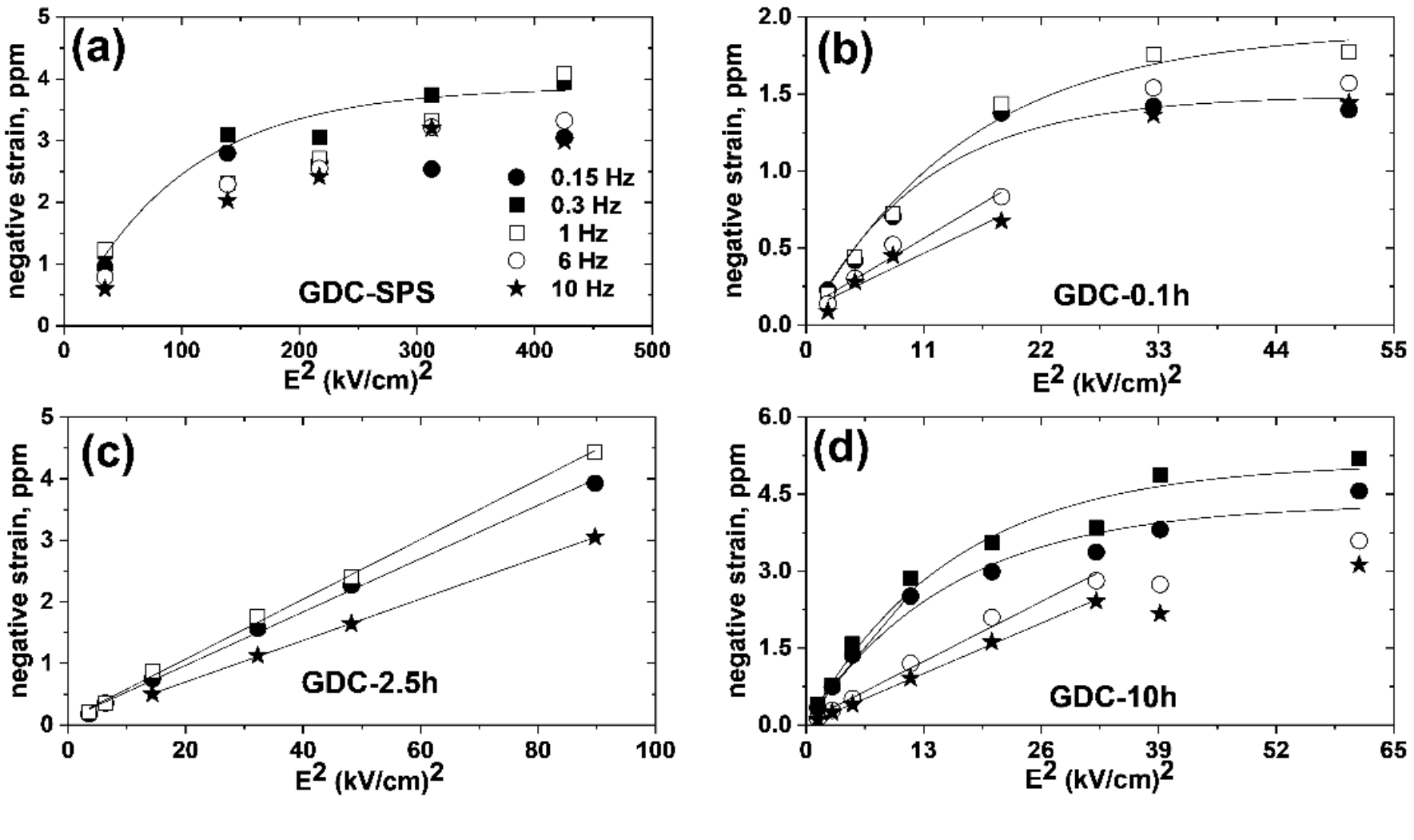

Figure 6: Electrostrictive negative strain as a function of applied electric field square at frequencies 0.15-10 Hz, showing strain saturation behavior at lower frequency.

Fig.7 demonstrates the declining trend of electrostrictive strain coefficient ($M_{33}$) with increasing frequency. This type of electrostriction relaxation with frequency can be fitted by (non-ideal-Debye) following function:

$$M_{33}\,(f) = \frac{M_{33}^{0}}{\sqrt{1+(\tau.f)^{2+\alpha}}} + M_{33}^{\infty} \tag{2}$$

Here, $M_{33}^{0}$ and $M_{33}^{\infty}$ are frequency independent electrostriction coefficient, $\tau$ is the relaxation time and $\alpha$ is detonated as non-ideality factor. Both the saturation and relaxation phenomena are observed in recent publication of Yavo *et al.* [12], specifying that both mechanisms are an intrinsic properties of electromechanical behavior of gadolinium doped ceria. $M_{33}$ values are approximately in order of $\sim\geq 10^{-18}$ $(m/V)^2$ at 10 Hz for all the samples, which are still one of order of magnitude higher than classical electrostriction model.

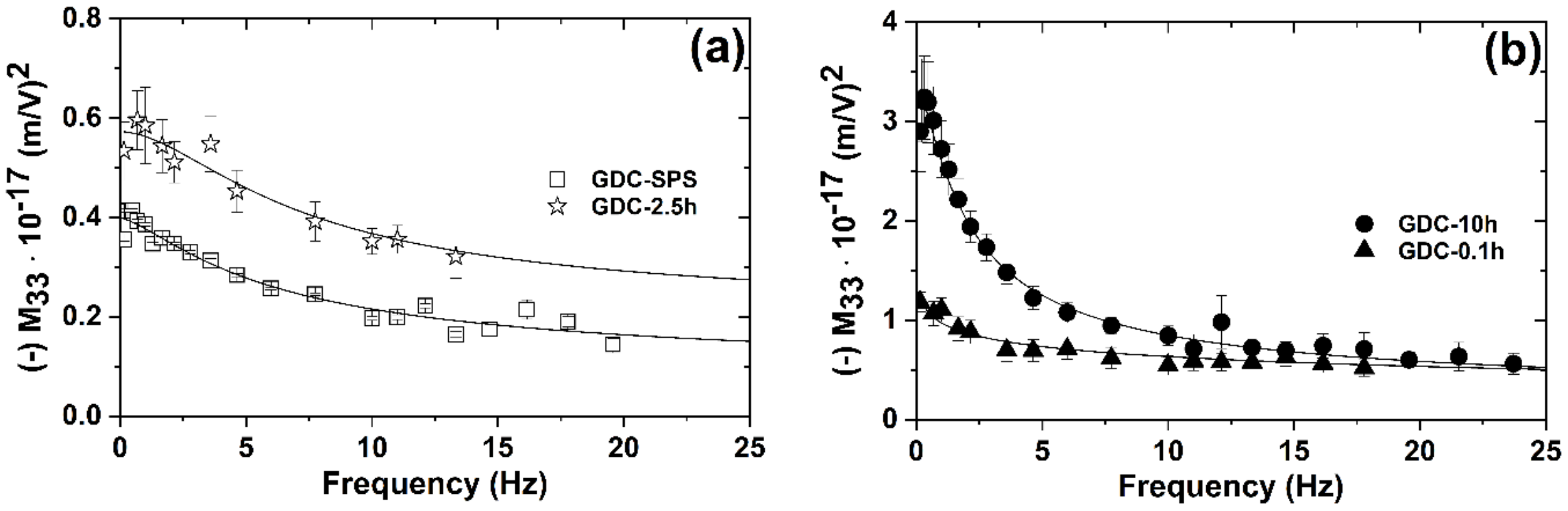

Figure 7: Frequency dependent relaxation of electrostriction strain coefficient ($M_{33}$), for conventional and non-conventional sintered samples, for frequencies 0< f< 25 Hz.

Table 2 illustrates the comparative analysis of total resistivity and electrostriction co-efficient among all the samples. It is shown that electrostriction increases with decreasing resistivity to a certain value (GDC-10h) then drops to a significant low level. The effect of blocking in electrostriction is also schematically presented in Table 2. The blocking diagram and the data strongly suggests that electrostriction is not dependent on the geometrical ratio between bulk and grain boundary. Microstructure does not necessarily influence the electromechanical properties. Additionally, it also confirms that the nominal oxygen vacancy concentration is not a true parameter that controls electrostriction. In conclusion, it is the blocking barrier at the grain boundaries, which regulates the electrostrictive properties. The blocking barrier is tuned by the configuration of oxygen vacancy within the grain boundary. Despite more uniform oxygen vacancy distribution, both GDC-SPS and GDC-0.1h materials show considerably limited electromechanical activity compared to GDC-10h with low density of large blocking barriers. These results finally conclude the dominant distribution of oxygen defect configuration to the electromechanical properties.

Table 2: Comparative analysis of resistivity and electrostriction coefficient between samples.

| Material Properties | Nano size grain | | Micron size grain | |
|---|---|---|---|---|
| | GDC-SPS | GDC-0.1h | GDC-10h | GDC-2.5h |
| Resistivity (≈ 300 °C) [Ω.cm] | ~100000 | ~50000 | ~30000 | ~7000 |
| Electrostriction (≈ 1 Hz) $(m/V)^2$ | $0.4 * 10^{-17}$ | $1.0 * 10^{-17}$ | $2.8 * 10^{-17}$ | $0.6 * 10^{-17}$ |

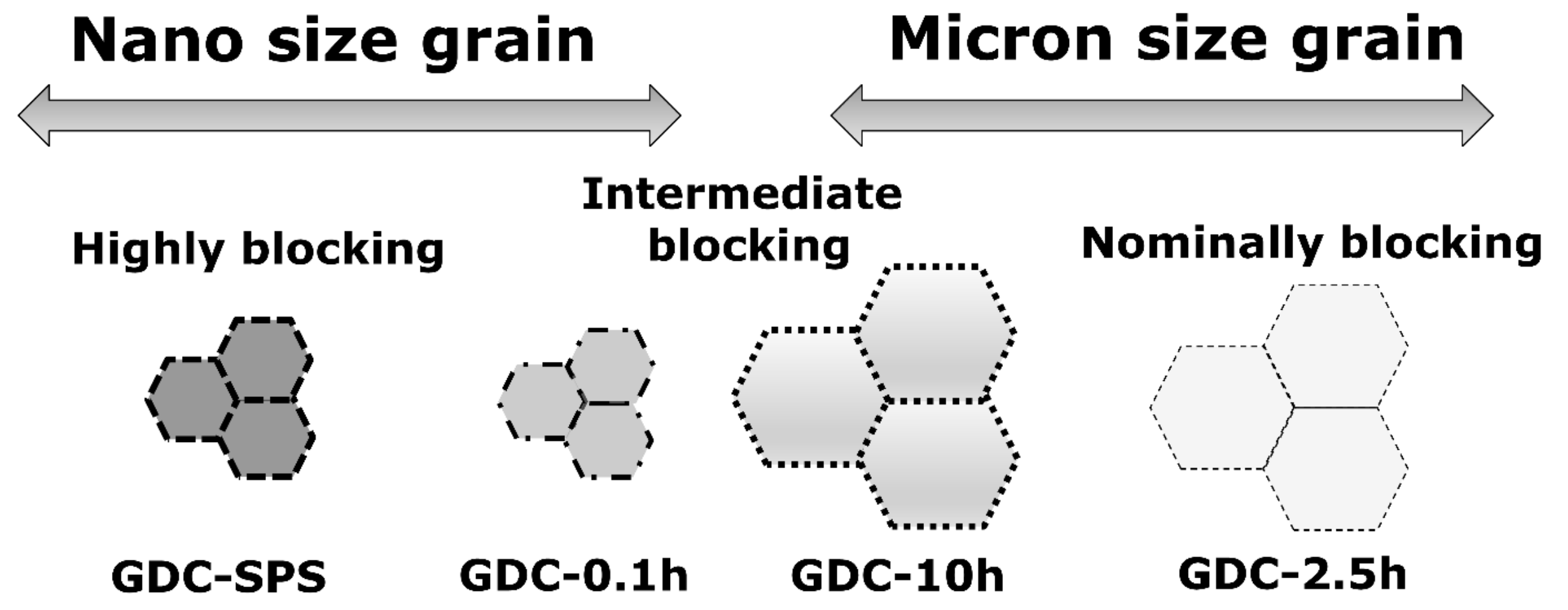

## 4. Conclusion

In this work, highly dense GDC ceramic pellets were fabricated by both non-conventional and conventional sintering methods. Non-conventional sintering was performed by SPS and fast firing to achieve similar nanometric microstructures with tuned oxygen vacancy configurations, with the same nominal oxygen vacancy concentration. The resulting polycrystalline materials exhibit unrelaxed microstructure with nano-grains, while the samples sintered in conventional method exhibits equilibrium grain of micron size. Surprisingly, electro-chemo-mechanical properties of the samples did not follow a mere geometrical grain size dependency. They show a strict dependency with ionic migration blocking barriers built in the materials by the different sintering processes. Furthermore, all the compounds show non-classical giant electrostriction with strong dependency on frequency and electric field amplitude. Above all, it was observed that sample with high bulk and low grain boundary relaxation frequency exhibits large electrostrictive coefficient, which is further related to the distribution of oxygen vacancies. In summary, the oxygen defects configuration rather than their nominal concentration in the bulk, controls the electromechanical behavior in Gd-doped ceria.

## 5. Acknowledgements

This research was supported by DFF-Research Project Grants from the Danish Council for Independent Research, Technology and Production Sciences, June 2016 and European H2020-FETOPEN-2016-2017 project BioWings (Partially), grant number 801267. The authors would like to thank Massimo Rosa for assistance in TEM.